\documentclass[superscriptaddress,nofootinbib,notitlepage]{revtex4-1}
\usepackage{amsfonts}
\usepackage[bookmarksopen,colorlinks]{hyperref}
\usepackage{graphicx}
\usepackage{dcolumn}
\usepackage{bm}
\usepackage{amssymb,amsmath}
\usepackage[dvipsnames]{xcolor}

\def\bf{\textbf}

\begin{document}

\title{Quest for the extra degree of freedom in $f(T)$ gravity}
\author{Rafael Ferraro}\affiliation{Instituto de Astronom\'ia y F\'isica
del Espacio (IAFE, CONICET-UBA), Casilla de Correo 67, Sucursal 28,
1428 Buenos Aires, Argentina.} \affiliation{Departamento de
F\'isica, Facultad de Ciencias Exactas y Naturales, Universidad de
Buenos Aires, Ciudad Universitaria, Pabell\'on I, 1428 Buenos Aires,
Argentina.} \email{ferraro@iafe.uba.ar}
\author{Mar\'ia Jos\'e Guzm\'an}\affiliation{Instituto de F\'isica de
La Plata (IFLP, CONICET-UNLP), C. C. 67, 1900 La Plata, Argentina.}
\email{mjguzman@fisica.unlp.edu.ar}

\begin{abstract}
{\ It has recently been shown that $f(T)$ gravity has $\frac{n(n-3)}{2}+1$
physical degrees of freedom (d.o.f.) in $n$ dimensions, contrary to previous
claims. The simplest physical interpretation of this fact is that the theory
possesses a scalar d.o.f. This is the case of $f(R)$ gravity, a theory that
can be understood in the Einstein frame as general relativity plus a
scalaron. The scalar field that represents the extra d.o.f.~in $f(T)$
gravity encodes information about the parallelization of the spacetime,
which is detected through a reinterpretation of the equations of motion in
both the teleparallel Jordan and Einstein frames. The trace of the equations
of motion in $f(T)$ gravity shows the propagation of the scalar d.o.f.,
giving an accurate proof of its existence. We also provide a simple toy
model of a physical system with rotational pseudoinvariance, like $f(T)$
gravity, which gives insights into the physical interpretation of the extra
d.o.f. We discuss some implications and unusual features of the previously
worked out Hamiltonian formalism for $f(T)$ gravity. Finally we show some
mathematical tools to implement the Hamiltonian formulation in the Einstein
frame of $f(T)$ gravity, which exhibits some problems that should be
addressed in future works. }
\end{abstract}

\maketitle

\section{Introduction}

Gravitational theories based on a spacetime with absolute parallelism are
extensions of general relativity (GR) that are being thoroughly studied in
the literature. They intend to solve the main problems of cosmology like the
hypothesis of dark matter, the accelerated expansion of the Universe and the
inflation paradigm, as well as theoretical problems of general relativity
like the emergence of singularities, and the quest for a quantum field
theory of gravity. Although teleparallel and modified teleparallel theories
of gravity have a long history, their applications in the realm of cosmology
started to being considered just recently in the context of teleparallelism
\textit{\`{a} la} Born-Infeld \cite{Ferraro:2006jd}. The original motivation
of this work was to obtain an early accelerated expansion of the Universe
without resorting to an inflaton field. This scheme addresses the problems
of general relativity at the high-energy regime, which has been studied and
discussed to some extent in subsequent work \cite%
{Ferraro:2008ey,Fiorini:2009ux,Ferraro:2009zk,
Ferraro:2010at,Fiorini:2013kba,Fiorini:2015hob,Fiorini:2016zrt}. The
article \cite{Ferraro:2006jd} is a particular case of the so-called
$f(T)$ gravity, a slight variation of teleparallel models that has
also been proposed as an alternative explanation of the late-time
accelerated expansion of the Universe \cite{Bengochea:2008gz}. Since
then, there has been a growing interest in this kind of theories
\cite{Wu:2010mn,Myrzakulov:2010vz,Wu:2010xk,Yang:2010hw,Bamba:2010iw,Yang:2010ji,
Chen:2010va,Bengochea:2010sg,Bamba:2010wb,Li:2010cg,Ferraro:2011us,
Sotiriou:2010mv,Li:2011wu,Li:2011rn,Ferraro:2011ks,Ferraro:2011zb,Ferraro:2012wp,
Bamba:2012vg,Izumi:2012qj,Fiorini:2013hva,Bamba:2013jqa,Nashed:2013mga,Chen:2014qtl,
Bejarano:2014bca,Ferraro:2014owa,
Wright:2016ayu,Farrugia:2016xcw,Nunes:2016qyp,
Bejarano:2017akj,Golovnev:2018red,Skugoreva:2017vde,
Hohmann:2017jao,Ferraro:2018tpu,Ong:2018heg,Golovnev:2018wbh} and
several other gravity models based in parallelized spacetimes have
been
developed \cite%
{Gonzalez:2015sha,Bahamonde:2016kba,Bahamonde:2016grb,Bahamonde:2017wwk,
Hohmann:2017duq,Nester:2017wau}. Modified teleparallel theories of gravity
on all its forms are expected to be an effective theory of a more
fundamental physical theory valid at energy regimes near the Planck energy.

There is one issue about $f(T)$ gravity that has given rise to disputes in
the literature: the fact that it is not invariant under local Lorentz
transformations of the tetrad field. This fact is usually interpreted as the
selection of preferred frames that parallelize the spacetime, which implies
that the theory contains extra degrees of freedom (d.o.f.)~if compared with GR. Although this
issue could be tackled by replacing the standard Weitzenb\"{o}ck connection
with a general one to obtain a covariant version of the theory \cite%
{Krssak:2015oua, Golovnev:2017dox,Hohmann:2018rwf}, the number of
d.o.f.~would remain unchanged since the same dynamical equations should be
expected in such covariant versions of $f(T)$ gravity.

A recent work has established that $f(T)$ gravity contains one extra
d.o.f.~compared with the teleparallel equivalent of general relativity (TEGR) in arbitrary dimension \cite{Ferraro:2018tpu}.
Although there could be many physical objects possessing a single degree of
freedom, the easiest interpretation would be that the theory is equivalent
to the teleparallel equivalent of general relativity plus a scalar field
minimally coupled to it. However, a very known fact is that $f(T)$ gravity
does not possess the analog of an Einstein frame, at least not in the way
that it occurs in $f(R)$ gravity \cite{Yang:2010ji,Wright:2016ayu}.
Therefore, it is timely to discuss the interpretation of the physical
degrees of freedom of $f(T)$ gravity in the light of its $f(R)$ counterpart.
We will analyze whether the scalar field interpretation is valid both in the
teleparallel Jordan and Einstein equivalent frames and look for further
insights in their respective Hamiltonian formulations.

This work is organized as follows. In Sec.~\ref{sec:fR} we will review
the basics of $f(R)$ gravity and the interpretation of the additional degree
of freedom on this theory, together with the definition of the Jordan and
Einstein frames. In Sec.~\ref{sec:tegrfT} we will introduce the TEGR and its simplest
generalization, $f(T)$ gravity. In Sec.~\ref{sec:extradof} we will
analyze the issue of the extra degree of freedom in $f(T)$ gravity through
the comparison of the teleparallel Jordan and the Einstein frames actions
and their equations of motion. In Sec.~\ref{sec:toymodel} we will present
a mechanical system with rotational pseudoinvariance, which is useful to
exemplify several uncommon features of $f(T)$ gravity and will help to
understand the physical meaning of the extra d.o.f. In Sec.~\ref{sec:HfT}
we will discuss some features of the Hamiltonian formalism for $f(T)$
gravity developed in \cite{Ferraro:2018tpu}. In Sec.~\ref{sec:HTEF} we
will introduce the Hamiltonian formalism for the teleparallel Einstein frame
and discuss the doability of the Dirac-Bergmann algorithm in two different
approaches. Finally we will dedicate Sec.~\ref{sec:concl} to the
conclusions.

\section{$f(R)$ gravity}

\label{sec:fR}

One of the most studied and simplest modifications of general relativity
consists in replacing the GR Lagrangian $R$ with a function $f(R)$ of the
scalar curvature. Such a gravitational theory is described by the action
\begin{equation}
S=-\dfrac{1}{2\kappa }\int d^{4}x\ \sqrt{-g}~f(R)+S_{matter},  \label{fRac}
\end{equation}
where $\kappa =8\pi G$. In \eqref{fRac} and throughout this work we adopt
the convention $(+,-,-,-)$ for the metric signature. The variation of the
action \eqref{fRac} gives rise to the following fourth-order dynamical
equations
\begin{equation}
f^{\prime }(R)~R_{\mu \nu }-\dfrac{1}{2}~f(R)~g_{\mu \nu }-[\nabla _{\mu
}\nabla _{\nu }-g_{\mu \nu }~\square ]f^{\prime }(R)=\kappa ~\mathcal{T}
_{\mu \nu },  
\label{eomfR}
\end{equation}
where $f^{\prime }(R)=df/dR$ and $\mathcal{T}_{\mu \nu }$ is the matter
stress-energy tensor. We see that $f^{\prime }(R)$ acts as a renormalization
of the gravitational constant $\kappa $, and therefore only functions satisfying
$f^{\prime }(R)>0$ should be considered; besides it should be $f^{\prime
\prime }(R)>0$ in order to avoid instabilities \cite%
{Dolgov:2003px,Nojiri:2003ft,Faraoni:2006sy,Olmo:2006eh}.

Taking the trace of the equations of motion \eqref{eomfR} we can see that
the Ricci scalar $R$ is \textit{dynamically} determined by the trace $%
\mathcal{T}$ of the energy-momentum tensor $\mathcal{T}_{\mu \nu }$ through
the second-order equation
\begin{equation}
f^{\prime }(R)~R-2~f(R)+3~\square f^{\prime }(R)=\kappa ~\mathcal{T}~
\label{fRtrace}
\end{equation}%
(in four dimensions). While Einstein's equations imply that $R$ vanishes in
vacuum, this equation possesses nontrivial propagating solutions in vacuum,
so suggesting the existence of a new degree of freedom related to $R$ [or $f^{\prime }(R)$]. The presence of additional degrees of freedom in $f(R)$
gravity is foreseeable, since the Cauchy data for the higher (fourth) order
of the differential equations involve more functions to be chosen on the
Cauchy surface. The issue of how many new degrees of freedom are involved in
$f(R)$ gravity can be better understood in the so-called Jordan frame, which
we proceed to review.

\subsection{Jordan frame}

Equations \eqref{eomfR} are fourth-order differential equations for the
metric. But, alternatively, one could regard the set \eqref{eomfR}, %
\eqref{fRtrace} as second-order equations for the metric and the scalar
object $f^{\prime }(R)$. Without any harm let us change the notation to
\begin{equation}
\phi \equiv f^{\prime }(R),\ \ \ V(\phi )\equiv R~\phi -f(R),
\label{LegTra}
\end{equation}
to rewrite Eqs. \eqref{eomfR} as \cite{Sotiriou:2008rp}
\begin{equation}
R_{\mu \nu }-\dfrac{1}{2}\,g_{\mu \nu }~R=\dfrac{\kappa }{\phi }~\mathcal{T}%
_{\mu \nu }-\dfrac{g_{\mu \nu }}{2\phi }~V(\phi )+\dfrac{1}{\phi }~[\nabla
_{\mu }\nabla _{\nu }\phi -g_{\mu \nu }\,\square \phi ].  \label{fReom-Jf}
\end{equation}%
Equation \eqref{LegTra} defines the Legendre transform of $f(R)$. Therefore
one also obtains
\begin{equation}
R=V^{\prime }(\phi ).  \label{Rjf}
\end{equation}
Equations \eqref{fReom-Jf} \ and \eqref{Rjf} can be recognized as the
dynamical equations associated with the action
\begin{equation}
S_{JF}[g_{\mu \nu },\phi ]=-\dfrac{1}{2\kappa }\int d^{4}x\ \sqrt{-g}~[\phi
~R-V(\phi )]+S_{matter}  \label{Sjf}
\end{equation}%
(which resembles a Brans-Dicke action with $\omega _{BD}=0$ and a
nontrivial potential). By varying the action \eqref{Sjf} with respect to $%
\phi $ one gets Eq.~\eqref{Rjf}, which says the scalar $\phi $ is a function
of the scalar curvature $R$. This result implies that the Lagrangian in %
\eqref{Sjf} is dynamically equivalent to the Legendre transform of the
function $V(\phi )$; then, it can be rewritten as a function of $R$
uniquely:
\begin{equation}
f(R)=\phi ~R-V(\phi ),  \label{InvLegTra}
\end{equation}
which completes the full circle of the equivalence between the actions %
\eqref{Sjf} and \eqref{fRac}. In fact, Eqs.~\eqref{LegTra}, \eqref{Rjf} and %
\eqref{InvLegTra} contain the entire mechanism of the Legendre transform and
its inverse ($f^{\prime \prime }\neq 0$ in order that the transformation be
invertible). The action \eqref{Sjf} is said to be the \textit{Jordan frame}
representation of $f(R)$ gravity.

Notice that one can use the result \eqref{Rjf} to write the trace of Eq.~%
\eqref{fReom-Jf} as a wave equation for a self-interacting scalar field $%
\phi $:
\begin{equation}
3~\square \phi -\phi ^{3}~[\phi ^{-2}~V(\phi )]^{\prime }=\kappa ~\mathcal{T}.  
\label{fRtrace-Jf}
\end{equation}
From this perspective, $\phi $ looks like a scalar field minimally coupled
to the metric $g_{\mu \nu }$, but coupled to the matter. On the other hand, Eqs.~\eqref{fReom-Jf} are Einstein equations sourced by the matter
and the scalar field. Thus, $f(R)$ gravity can be rephrased as a
scalar-tensor theory with second-order dynamical equations \cite%
{Barrow:1988xh,Maeda:1988ab,Chiba:2003ir}. As dynamical equations for the
metric $g_{\mu \nu }$, Eqs.~\eqref{fReom-Jf} keep all the
symmetries contained in Einstein equations \footnote{%
In particular, the automatic conservation of $\mathcal{T}_{\mu \nu }$ is
guaranteed.}; so, they describe two d.o.f.~in four dimensions. Besides,
there is one extra d.o.f.~described by the second-order equation %
\eqref{fRtrace-Jf}.

\subsection{Einstein frame}

An even clearer understanding of the extra scalar degree of freedom in $f(R)$
gravity is achieved by working in the so-called \textit{Einstein frame}. In
this representation the extra d.o.f.~is distilled in the action as a scalar
field possessing its own kinetic term. This goal is attained by means of a
conformal transformation of the metric together with a redefinition of the
scalar field:
\begin{equation}
g_{\mu \nu }\longrightarrow \tilde{g}_{\mu \nu }=\phi ~g_{\mu \nu },\ \ \ \ \ \phi \longrightarrow \tilde{\phi}=\sqrt{\dfrac{3}{2\kappa }}\ln
\phi,  
\label{conf}
\end{equation}
(then it is $\sqrt{-g}=\phi ^{-2}\sqrt{-\tilde{g}}$). Thus, applying the
relation for the scalar curvatures of conformally related metrics
\begin{equation}
\phi ~\tilde{R}=R-\dfrac{3}{2}~g^{\mu \nu }~\partial _{\mu }\ln \phi
~\partial _{\nu }\ln \phi -3~\square \ln \phi,
\end{equation}
the action $S_{JF}$ in \eqref{Sjf} is rewritten as
\begin{equation}
S_{EF}[\tilde{g}_{\mu \nu },\tilde{\phi}]=\int d^{4}x\ \sqrt{-\tilde{g}}%
~\left( -\dfrac{\tilde{R}}{2\kappa }+\dfrac{1}{2}~\tilde{g}^{\mu \nu
}~\partial _{\mu }\tilde{\phi}~\partial _{\nu }\tilde{\phi}-U(\tilde{\phi}
)\right) +S_{matter},  
\label{Sef}
\end{equation}
where the new potential $U(\tilde{\phi})$ is
\begin{equation}
U(\tilde{\phi})=-\dfrac{V(\phi )}{2\kappa ~\phi ^{2}}=\dfrac{f(R)-R~f^{\prime }(R)}{2\kappa ~f^{\prime }(R)^{2}}.
\end{equation}%
The action \eqref{Sef} describes a tensor field $\tilde{g}_{\mu \nu }$
governed by Einstein equations and a minimally coupled scalar field $\tilde{\phi}$ governed by a Klein-Gordon equation that includes a self-interaction
\cite{Faraoni:1999hp,Bhadra:2006rn,Capozziello:2010sc,Sk.:2016tvl}.
Therefore, the Einstein frame has the virtue of explicitly showing the extra
d.o.f.~at the level of the action, since $S_{EF}$ is the sum of the
Einstein-Hilbert action and the scalar field action. Of course $\tilde{g}_{\mu \nu }$ and $\tilde{\phi}$ are not the original tensor and scalar fields;
but it does not matter for the purpose of counting degrees of freedom.
Anyway, once the Einstein frame dynamical equations are solved, one can
always return to the original metric $g_{\mu \nu }$ by means of Eq.~\eqref{conf} .

Our very aim is to investigate whether the issue of the extra d.o.f.~in $%
f(T) $ gravity can be addressed in analogy with the $f(R)$ gravity case. For
this, we will introduce $f(T)$ theories of gravity and its progenitor, the
teleparallel equivalent of general relativity.

\section{Teleparallel and $f(T)$ gravity}

\label{sec:tegrfT}

\subsection{Teleparallel gravity}

In order to build a dynamical theory for a field of orthonormal frames, we
begin by defining a manifold $M$ and a basis of vectors $\{\mathbf{e}_{a}\}$
in the tangent space $T_{p}(M)$. We also define a dual basis $\{\mathbf{E}%
^{a}\}$ in the cotangent space $T_{p}^{\ast }(M)$ such that $\mathbf{E}^{a}(%
\mathbf{e}_{b})=\delta _{b}^{a}$. If expanded in a coordinate basis,
\begin{equation}
\mathbf{e}_{a}=e_{a}^{\mu }~\partial _{\mu },\ \ \ \mathbf{E}^{a}=E_{\mu
}^{a}~dx^{\mu },
\end{equation}%
the duality relationships are written as
\begin{equation}
E_{\mu }^{a}~e_{b}^{\mu }=\delta _{b}^{a},\ \ \ e_{a}^{\mu }~E_{\nu
}^{a}=\delta _{\nu }^{\mu }.  \label{duality}
\end{equation}%
We will use Greek letters $\mu ,\nu ,\ldots =0,\ldots ,n-1$ for spacetime
coordinate indices and Latin letters $a,b,\ldots =0,\ldots ,n-1$ for
tangent space or Lorentz indices. A \textit{vielbein} is a basis that
encodes the metric structure through the following relation:
\begin{equation}
\mathbf{g}=\eta _{ab}\ \mathbf{E}^{a}\otimes \mathbf{E}^{b};  \label{orth1}
\end{equation}
then the inverse expression
\begin{equation}
\mathbf{e}_{a}\cdot \mathbf{e}_{b}=\mathbf{g}(\mathbf{e}_{a}, \mathbf{e}_{b}) = \eta_{ab}  \label{orth2}
\end{equation}
states that the vielbein is an orthonormal basis. In $n=4$ we have a \textit{vierbein} or tetrad. The expressions \eqref{orth1} and \eqref{orth2} can be
alternatively written in coordinate language as
\begin{equation}
g_{\mu \nu }=\eta_{ab} E_{\mu }^{a} E_{\nu }^{b},\ \ \ \eta_{ab} = g_{\mu \nu } e_{a}^{\mu } e_{b}^{\nu }.  
\label{metric}
\end{equation}
This notation also implies that $\sqrt{-g}=\text{det}(E_{\mu }^{a})\equiv E$%
, and also $e\equiv \text{det}(e_{a}^{\mu })=E^{-1}$. We can construct a
dynamical theory for the spacetime through the tetrad field, since it
encodes the metric tensor and henceforth the geometry of the spacetime. To
get equations of motion equivalent to Einstein equations we use the TEGR
action,
\begin{equation}
S=\dfrac{1}{2\kappa }\int d^{4}x\ E~T+S_{matter},  \label{ltegr}
\end{equation}
where $T$ is a local Lorentz pseudoinvariant made up of the torsion of the
Weitzenb\"{o}ck connection
\begin{equation}
T_{\ \nu \rho }^{\mu }=e_{a}^{\mu }~(\partial _{\nu }E_{\rho }^{a}-\partial
_{\rho }E_{\nu }^{a})~.  \label{torsion}
\end{equation}%
In terms of this torsion, $T$ is written as
\begin{equation}
T\equiv T_{\ \mu \nu }^{\rho }~S_{\rho }^{\ \mu \nu },
\label{pseudoinvariant}
\end{equation}%
where the object $S_{\rho }^{\ \mu \nu }$ is called the superpotential and
is given by
\begin{equation}
S_{\rho }^{\ \mu \nu }\equiv -\dfrac{1}{4}~(T_{\ \ \rho }^{\mu \nu }-T_{\ \
\rho }^{\nu \mu }-T_{\rho }^{\ \ \mu \nu })+\dfrac{1}{2}~T_{\sigma }^{\
\sigma \mu }~\delta _{\rho }^{\nu }-\dfrac{1}{2}~T_{\sigma }^{\ \sigma \nu
}~\delta _{\rho }^{\mu }.  
\label{Spot}
\end{equation}
GR and TEGR are equivalent theories because the GR Lagrangian $R$ differs
from the TEGR Lagrangian in a boundary term,
\begin{equation}
R=-T+2~e~\partial _{\mu }(E~T^{\mu }),  \label{Lagreq}
\end{equation}
where $R$ is the Ricci scalar of the Levi-Civita connection associated with
the metric \eqref{metric} and $T^{\mu }\equiv T_{\ \rho }^{\rho \ \mu }$ is
the vector part of the torsion.

By varying the action \eqref{ltegr} with respect to the tetrad field we
obtain the following dynamical equations:
\begin{equation}
4~e~\partial _{\mu }(E~e_{a}^{\lambda }~S_{\lambda }^{\ \mu \nu
})+4~e_{a}^{\lambda }~T_{\ \mu \lambda }^{\rho }~S_{\rho }^{\ \mu \nu
}-e_{a}^{\nu }~T=-2\kappa ~e_{a}^{\lambda }~\mathcal{T}_{\lambda }^{\ \nu },
\label{eomtegr}
\end{equation}
which prove to be equivalent to Einstein equations for the metric~\eqref{metric}. In particular, the trace of \eqref{eomtegr} in four
dimensions is%
\begin{equation}
2~e~E_{\nu }^{a}~\partial _{\mu }(E~e_{a}^{\lambda }~S_{\lambda }^{\ \mu \nu
})=-\kappa ~\mathcal{T},  
\label{Teom}
\end{equation}
where $\mathcal{T}$ is the trace of the energy-momentum tensor. However it
can be proven that
\begin{equation}
2~e~E_{\nu }^{a}~\partial _{\mu }(E~e_{a}^{\lambda }~S_{\lambda }^{\ \mu \nu
})=-T+2~e~\partial _{\mu }(E~T^{\mu }).  
\label{id4div}
\end{equation}
Therefore it results
\begin{equation}
-T+2~e~\partial _{\mu }(E~T^{\mu })=-\kappa ~\mathcal{T}.  
\label{R}
\end{equation}
The lhs is $R$, as we know from the Lagrangian equivalence \eqref{Lagreq}.
Thus we retrieve the trace of Einstein equations.

\subsection{\textit{f}(\textit{T}) gravity}

The teleparallel equivalent of general relativity can be utilized as an
underlying framework for a more general class of modified gravities. The
simplest choice is to generalize the TEGR Lagrangian through an arbitrary
function of the torsion scalar $T$; this gives rise to the so-called $f(T)$
gravity. The action for this theory is given by
\begin{equation}
S[E_{\mu }^{a}]=\dfrac{1}{2\kappa }\int d^{4}x\ E\ f(T)+S_{matter},
\label{fTact}
\end{equation}
where $T$ is the Weitzenb\"{o}ck or torsion scalar, which is quadratic in
first-order derivatives of the tetrad field. This fact alone implies that
the equations of motion of this theory will always be second order, which is
an advantage over other modified gravities [as metric $f(R)$ gravity]. In
fact, by varying the action \eqref{fTact} with respect to the tetrad, it
results \cite{Ferraro:2012wp}
\begin{equation}
4~e~\partial _{\mu }(f^{\prime }(T)\,E~e_{a}^{\lambda }~S_{\lambda }^{\ \mu
\nu })+4~f^{\prime }(T)\,e_{a}^{\lambda }~T_{\ \mu \lambda }^{\sigma
}~S_{\sigma }^{\ \mu \nu }-e_{a}^{\nu }~f(T)=-2\kappa ~e_{a}^{\lambda }~%
\mathcal{T}_{\lambda }^{\ \nu }.  
\label{eomfT}
\end{equation}
Equation \eqref{eomfT} shows that $f^{\prime }(T)$ renormalizes the
gravitational constant $\kappa $ and the volume $E$. It has been known since
the beginning of the theory that the action of $f(T)$ gravity and their
equations of motion are sensitive to local Lorentz transformations in the
tangent space of the spacetime manifold: given a tetrad $\mathbf{E}^{a}$
that solves Eqs.~\eqref{eomfT}, the Lorentz transformed tetrad $\mathbf{E}^{a^{\prime }}=\Lambda _{\ a}^{a^{\prime }}(x) \mathbf{E}^{a}$
will not necessarily solve them. The departure from the local Lorentz
invariance can be already recognized in the action; the torsion scalar $T$
is invariant under \textit{global} Lorentz transformations of the tetrad,
but local transformations will provide it with terms containing derivatives
of the Lorentz matrices $\Lambda _{\ a}^{a^{\prime }}(x)$. \ For TEGR this
fact is harmless; according to Eq.~\eqref{Lagreq}, $T$ is equal to $-R$
--which depends only on the (locally invariant) metric-- plus a boundary
term which is the responsible for the lack of local invariance. But a
boundary term is irrelevant for the dynamics in the TEGR action \eqref{ltegr}. On the contrary, the most general functional form $f$ \ in \eqref{fTact}
is not linear; thus the four-divergence term remains encapsulated in $f$, and the local Lorentz invariance is definitely lost.

\subsection{The premetric formulation of TEGR}

We will find useful in later sections and future work to rewrite several
mathematical objects by getting rid of the metric tensor in favor of the
tetrad field and the Minkowski metric. As it was firstly introduced in \cite{Ferraro:2016wht}, the TEGR Lagrangian can be written in a compact form like
\begin{equation}
L=E\ T=\frac{1}{2}~E\ \partial _{\mu }E_{\,\,\nu }^{a}\ \partial _{\rho
}E_{\,\,\lambda }^{b}\ e_{c}^{\mu }~e_{e}^{\nu }~e_{d}^{\rho
}~e_{f}^{\lambda }\ M_{ab}^{\ \ cedf},  \label{LagrM}
\end{equation}
where $M_{ab}^{\ \ cedf}$ is the \emph{supermetric}, the Lorentz invariant
tensor given by
\begin{equation}
M_{ab}^{\ \ cedf}=2~\eta_{ab}~\eta^{c[d}\eta^{f]e}-4~\delta_{a}^{[d}\eta^{f][c} \delta_{b}^{e]} + 8~\delta_{a}^{[c} \eta^{e][d} \delta_{b}^{f]}.
\label{supermetric}
\end{equation}
This object is nothing more than a \textit{constitutive tensor}, that is the
object that relates the excitation field and the field strength of any
physical theory \cite{Itin:2016nxk,Itin:2018dru}. This result can be derived
alternatively by writing both $T_{\ \mu \nu }^{\rho }$ and $S_{\rho }^{\ \mu
\nu }$ in the premetric formalism. Observing that
\begin{equation}
T_{\ \mu \nu }^{\rho }=e_{a}^{\rho }~(\partial _{\mu }E_{\nu }^{a}-\partial
_{\nu }E_{\mu }^{a})=\partial _{\sigma }E_{\lambda }^{a}~e_{a}^{\rho
}~e_{b}^{\sigma }~e_{c}^{\lambda }~E_{\mu }^{d}~E_{\nu }^{e}~(\delta
_{d}^{b}~\delta _{e}^{c}-\delta _{e}^{b}~\delta _{d}^{c}),
\label{TorsionPF}
\end{equation}
we can rewrite the superpotential $S_{\rho }^{\ \mu \nu }$ defined in \eqref{Spot} as
\begin{equation}
S_{\rho }^{\ \mu \nu }=E_{\rho }^{d}~\partial _{\sigma }E_{\lambda
}^{a}~e_{b}^{\sigma }~e_{c}^{\lambda }~e_{e}^{\mu }~e_{f}^{\nu }~S_{ad}^{\ \
becf},
\end{equation}
where we have defined
\begin{equation}
S_{ad}^{\ \ becf} = \eta_{ad}~\eta^{e[b} \eta^{c]f}-\eta^{f[b} \delta_{d}^{c]}~\delta_{a}^{e} + \eta^{e[b} \delta_{d}^{c]}~\delta_{a}^{f}+ 2~\delta_{a}^{b}~\eta^{c[e} \delta_{d}^{f]}-2~\delta_{a}^{c}~\eta^{b[e} \delta_{d}^{f]}.  \label{SpotPF}
\end{equation}
Performing the corresponding algebra, it is not difficult to check that, up
to a factor $E$, the multiplication of \eqref{TorsionPF} and \eqref{SpotPF}
gives \eqref{LagrM}. This is in virtue of the simple result
\begin{equation}
S_{ba}^{\ \ dcfe}-S_{ba}^{\ \ defc}=M_{ab}^{\ \ cedf}.
\end{equation}
We will need for later to rewrite the four-divergence $\partial_{\mu}(E T^{\mu })$, that distinguishes the (Levi-Civita) curvature scalar from
the torsion scalar. This is written as
\begin{equation}
\partial _{\mu }(E~T^{\mu })=\partial _{\mu }(E~g^{\mu \nu }e_{a}^{\rho
}~(\partial _{\rho }E_{\nu }^{a}-\partial _{\nu }E_{\rho }^{a})).
\label{diver}
\end{equation}
Taking into account that $\partial_{\mu }E = E e_{a}^{\lambda } \partial_{\mu } E_{\lambda}^{a}$, and replacing the metric tensor by its tetrad
counterpart, in the end it is obtained
\begin{equation}
B\equiv e~\partial _{\mu }(E~T^{\mu })=\dfrac{1}{2}\partial _{\mu }E_{\nu
}^{a}~\partial _{\rho }E_{\lambda }^{b}\ e_{c}^{\mu }~e_{e}^{\nu
}~e_{d}^{\rho }~e_{f}^{\lambda }\ B_{ab}^{\ \ cedf}+\partial _{\mu }\partial
_{\rho }E_{\nu }^{a}\ e_{b}^{\mu }e_{c}^{\nu }e_{d}^{\rho }\ (\delta
_{a}^{d}\eta ^{bc}-\delta _{a}^{c}\eta ^{bd}),  
\label{Bterm}
\end{equation}
where the analog of the constitutive tensor for the four-divergence is
written as
\begin{equation}
B_{ab}^{\ \ cedf} = 4~\eta^{cd}~\delta_{a}^{[f}\delta_{b}^{e]} + 4~\eta^{ce}~\delta_{a}^{[d} \delta_{b}^{f]} + 4~\eta^{cf}~\delta_{a}^{[e} \delta_{b}^{d]} + 4~\delta_{a}^{c}~\eta^{e[d} \delta_{b}^{f]}.
\end{equation}
Remarkably this object does not have the same antisymmetry properties as $M_{ab}^{\ \ cedf}$.

\subsection{Approaches on the degrees of freedom of \textit{f}(\textit{T})
gravity}

While in $f(R)$ gravity the increase of the number of degrees of freedom
manifests itself through the increase of the order of the differential
equations, in $f(T)$ gravity the dynamical equations remain as second-order
equations. Instead, the $f(T)$ dynamical equations have less gauge freedom
than the TEGR equations, because not all the local Lorentz transformations
of the tetrad are symmetries of the dynamics. Less gauge freedom implies
more genuine degrees of freedom.

There are several approaches to the question of the number and nature of the
degrees of freedom of $f(T)$ gravity, which do not necessarily coincide with
their outcomes. The Hamiltonian analysis of $f(T)$ gravity presented in \cite{Li:2011rn}, based on Maluf's Hamiltonian formulation of TEGR \cite{Maluf:2001rg}, gives $n-1$ extra d.o.f.~for $f(T)$ gravity in arbitrary
dimension. In \cite{Chen:2014qtl} it is performed a simple Hamiltonian
analysis based on the number of pairs of Lorentz constraints that would
become second-class. The authors establish that $f(T)$ gravity might have five, four or two  d.o.f.~based on the possibility that six, four or two Lorentz
constraints become second-class. The enforcement of the vanishing of the
Riemann tensor through the introduction of Lagrange multipliers has been
analyzed with detail in \cite{Nester:2017wau}. This strategy attempts to
avoid a particular choice of the spin connection but adds a large quantity
of Lagrange multipliers that are nevertheless determined at the end of the
procedure. Other approaches that need further development in order to give a
characterization of the degrees of freedom are the study of the remnant
group of local Lorentz symmetries \cite{Ferraro:2014owa}, the null tetrad
approach \cite{Bejarano:2014bca,Bejarano:2017akj} and the discussion on the
covariantization of the theory \cite{Krssak:2015oua,Golovnev:2017dox,Hohmann:2018rwf}, among others. Finally,
the recent work performed in \cite{Ferraro:2018tpu} suggests that there
exists only one additional degree of freedom in arbitrary dimension. The
counting of d.o.f. obtained in this work is not the same as the one
presented in \cite{Li:2011rn}, nor falls in any of the categories
sketched in \cite{Chen:2014qtl}. However, if the theory has one rightful
extra d.o.f., it could be similar to the scalaron that appears in $f(R)$
theories of gravity. We will discuss the possibility of finding a scalar
d.o.f.~in the equations of motion of $f(T)$ gravity in the next section.

\section{The extra degree of freedom of $f(T)$ gravity}

\label{sec:extradof}

\subsection{Jordan frame}

The previous discussions, and the comparison with $f(R)$ gravity in Sec.~\ref{sec:fR}, could suggest taking a look at the Jordan action as a way to
understand the nature of the extra degree of freedom. This approach is
successful in $f(R)$ gravity because it decouples the equations at the same
time that it reduces their differential order. However, in $f(T)$ gravity
the equations of motion \eqref{eomfT} are already second order. Besides, by
tracing them we do not succeed in isolating the dynamics of some scalar
degree of freedom; in fact, in four dimensions the trace is
\begin{equation}
e~E_{\nu }^{a}~\partial _{\mu }(f^{\prime }(T)\,E~e_{a}^{\lambda
}~S_{\lambda }^{\ \mu \nu })+f^{\prime }(T)~T-f(T)=-\frac{\kappa }{2}~
\mathcal{T},
\end{equation}
that can be rewritten as
\begin{equation}
e~\partial_{\mu }(f^{\prime }(T)\ E E_{\nu }^{a}~e_{a}^{\lambda}~S_{\lambda }^{\ \mu \nu })-f^{\prime }(T)\ e_{a}^{\lambda }~S_{\lambda}^{\ \mu \nu }\partial_{\mu } E_{\nu }^{a} + f^{\prime }(T)~T-f(T)=-\frac{\kappa }{2}~\mathcal{T}.
\end{equation}

Let us now consider the dynamics in the Jordan frame, which is defined by
the Legendre transform
\begin{equation}
\phi =f^{\prime }(T),\ \ \ V(\phi )=T~\phi -f(T),  \label{transform}
\end{equation}
also implying that it is
\begin{equation}
T=V^{\prime }(\phi ).  
\label{eomJ2}
\end{equation}
Thus the dynamical equations \eqref{eomfT} read \footnote{These equations have a nontrivial antisymmetric part $2 S_{[\rho \ \nu]}^{\ \ \mu} \partial_{\mu} \phi = -\kappa \mathcal{T}_{[\rho\nu]}$, where the rhs could be different from zero if the matter were coupled in a nontrivial way to the tetrad field.  In Ref. \cite{Golovnev:2017dox} the lhs has been obtained from another procedure, where the action is varied with respect to an ``inertial'' spin connection. This procedure has been used to find the spin connection in cosmological and spherically symmetric spacetimes in \cite{Hohmann:2018rwf}. }
\begin{equation}
4~\phi ^{-1}e~\partial _{\mu }(\phi \,E~e_{a}^{\lambda }~S_{\lambda }^{\ \mu
\nu })+4\,e_{a}^{\lambda }~T_{\ \mu \lambda }^{\sigma }~S_{\sigma }^{\ \mu
\nu }-e_{a}^{\nu }~T=-\frac{2\kappa }{\phi }~e_{a}^{\lambda }~\mathcal{T}_{\lambda }^{\ \nu }~+e_{a}^{\nu }~\frac{V(\phi )}{\phi },  
\label{eomJ1}
\end{equation}
which keep the structure of TEGR field equations, except for the
renormalizations of the gravitational constant $\kappa $ and the volume $E$,
and the term $\phi^{-1}V(\phi )$ playing the role of a local cosmological
constant. Equations \eqref{eomJ1} and \eqref{eomJ2} can be obtained by
varying the action
\begin{equation}
S_{JF}[E_{\mu }^{a},\phi ]=\dfrac{1}{2\kappa }\int d^{4}x~E~[\phi ~T-V(\phi
)]+S_{matter}.  
\label{actionJ}
\end{equation}
In particular, Eq.~\eqref{eomJ2} results from varying with respect
to $\phi $, which means that the Lagrangian is dynamically equivalent to the
Legendre transform of $V(\phi )$, and so it is a function $f(T)$ such that $\phi = f^{\prime }(T)$. Thus, the action \eqref{actionJ} is dynamically
equivalent to the action \eqref{fTact}.

Let us now analyze the trace of the dynamical equations. In four dimensions
it results that $S_{\nu }^{\ \mu \nu }=T_{\nu }^{\ \nu \mu }=T^{\mu }$.
Besides, it is $e_{a}^{\lambda } S_{\lambda }^{\ \mu \nu }\partial_{\mu} E_{\nu }^{a} = e_{a}^{\lambda } S_{\lambda }^{\ \mu \nu } \partial_{\lbrack
\mu } E_{\nu ]}^{a} = S_{\lambda }^{\ \mu \nu }T_{~~\mu \nu }^{\lambda }/2=T/2$. Then the equation for the trace becomes
\begin{equation}
2~e~\partial _{\mu }(f^{\prime }(T)\,E~T^{\mu })+f^{\prime
}(T)\,T-2~f(T)=-\kappa ~\mathcal{T}.  
\label{fpt}
\end{equation}
i.e.,
\begin{equation}
2~T^{\mu }\partial _{\mu }f^{\prime }(T)+2~f^{\prime }(T)\,e~\partial _{\mu
}(E~T^{\mu })+f^{\prime }(T)\,T-2~f(T)=-\kappa ~\mathcal{T}.
\end{equation}
From Eq.~\eqref{Lagreq}, $2 e \partial_{\mu }(E T^{\mu })$ can be
substituted with $R+T$. By writing Eq.~\eqref{fpt} in terms of $\phi $ one
gets
\begin{equation}
2~T^{\mu }\partial_{\mu }\phi +2~V(\phi )+\phi R = -\kappa \mathcal{T}.
\label{newmode2}
\end{equation}
In TEGR it is $\phi =1$ and $V(\phi)=0$, and the former equation just implies $R=-\kappa ~
\mathcal{T}~$ [cf. Eq.~\eqref{R}]. So, in principle, Eq.~\eqref{newmode2}
could be interpreted as describing the propagation of a mode that was not
present in TEGR, like Eq.~\eqref{fRtrace} in $f(R)$ gravity. The existence
of an extra degree of freedom is connected to the loss of a gauge symmetry.
In fact, TEGR dynamics is invariant under local Lorentz transformations of
the tetrad because the Lagrangian $T$ in \eqref{ltegr} differs from $R$ in a
four-divergence [see Eq.~\eqref{Lagreq}], $R$ being invariant under local
Lorentz transformations of the tetrad field since it only depends on the
metric. Thus TEGR is not a dynamical theory for the tetrad but for the
metric. Instead the gauge symmetries of the action \eqref{fTact} are reduced
to those of $T$, which does not remain invariant under a general local
Lorentz transformation. Such a remnant gauge symmetry of $f(T)$ gravity has
an on-shell character. Notice that the invariance of $T$ implies the
invariance of $\phi $ in Eq.~\eqref{eomJ2}. On the other hand, according to
Eq.~\eqref{Lagreq}, the invariance of $T$ is equivalent to the invariance of
the four-divergence $e \partial_{\mu }(E T^{\mu })$. Under local Lorentz
transformations of the tetrad field, $e \partial_{\mu }(E T^{\mu })$
transforms as
\begin{eqnarray}
e~\partial _{\mu }(E~T^{\mu }) &\longrightarrow &e~\partial _{\mu }(E~g^{\mu
\nu }\Lambda _{~a^{\prime }}^{a}e_{a}^{\rho }(\partial _{\rho }(\Lambda
_{~b}^{a^{\prime }}E_{\nu }^{b})-\partial _{\nu }(\Lambda _{~b}^{a^{\prime
}}E_{\rho }^{b})))  \notag \\
&=&e~\partial _{\mu }(E~T^{\mu })+e~\partial _{\mu }(E~g^{\mu \nu }(E_{\nu
}^{b}~e_{a^{\prime }}^{\rho }~\partial _{\rho }\Lambda _{~b}^{a^{\prime
}}-\Lambda _{~a^{\prime }}^{a}\partial _{\nu }\Lambda _{~a}^{a^{\prime }}));
\end{eqnarray}
for infinitesimal transformations
\begin{equation}
\Lambda_{\ \ b}^{a^{\prime} }=\delta_{\ \ b}^{a^{\prime}} - \frac{1}{2}\
\sigma^{gh}(x)\ (M_{gh})_{\ \ b}^{a^{\prime}}\ +\mathit{O}(\sigma^{2}),
\end{equation}
where the generators $M_{gh}$ are the traceless matrices $(M_{gh})_{\ \
b}^{a^{\prime} } = \delta_{g}^{a^{\prime} }~\eta_{hb}-\delta _{h}^{a^{\prime} }~\eta_{gb}$, it results
\begin{equation}
e~\partial _{\mu }(E~T^{\mu })\longrightarrow e~\partial _{\mu }(E~T^{\mu }) -
\frac{1}{2}~e~\partial _{\mu }(E~g^{\mu \nu }E_{\nu }^{b}~(e_{g}^{\rho
}~\eta _{hb}-e_{h}^{\rho }~\eta _{gb})~\partial _{\rho }\sigma ^{gh})+
\mathit{O}(\sigma ^{2}).
\end{equation}
By replacing $g^{\mu \nu }E_{\nu }^{b}~\eta_{hb}=e_{h}^{\mu }$, and
noticing that $(e_{g}^{\rho }~e_{h}^{\mu }-e_{g}^{\mu }~e_{h}^{\rho })$ $\partial_{\mu }\partial_{\rho }\sigma^{gh}=0$, we obtain that the remnant
gauge symmetry is generated by those Lorentz transformations that fulfill
[cf. Eq.~(32) in Ref.~\cite{Ferraro:2014owa}] 
\begin{equation}
\partial _{\mu }(E~(e_{g}^{\rho }~e_{h}^{\mu }-e_{g}^{\mu }~e_{h}^{\rho
}))~\partial_{\rho }\sigma ^{gh}=0.
\end{equation}
To finish this subsection, we will notice that Eq.~\eqref{newmode2} could
admit solutions $\phi =$ constant [i.e., $T=$ const, according to Eq.~\eqref{eomJ2}]. In such a case Eqs.~\eqref{eomJ1} look like TEGR
equations with peculiar gravitational and cosmological constants. Remarkably
more than one solution of this type could be found for the same potential $V(\phi )$ [i.e., for the same function $f(T)$] \cite{Skugoreva:2017vde}.

\subsection{Einstein frame}

Let us examine whether the issue of the number of degrees of freedom in $f(T) $ gravity can be better understood in the so-called Einstein frame. We prevent the reader from possible confusion clarifying that in this work we will refer to the "Einstein frame" representation of $f(T)$ gravity as the action that results from applying a conformal transformation to the Jordan frame action of $f(T)$ gravity. This is a somehow forced terminology, since it is common knowledge that the outcome of such conformal transformation is not the TEGR action plus a scalar field, as it would be desired given that a similar fact happens in $f(R)$ gravity.  We follow the standard procedure and start from the Jordan frame action \eqref{actionJ}, which could be regarded as the
particular case $\omega_{BD}=0$ of a sort of Brans-Dicke teleparallel
action of the form
\begin{equation}
S_{BD}=\dfrac{1}{2\kappa }\int d^{4}x~E~[\phi ~T+\omega _{BD}~\phi
^{-1}~g^{\mu \nu }~\partial _{\mu }\phi ~\partial _{\nu }\phi -V(\phi
)]+S_{matter}.
\end{equation}
Now we will consider a change of dynamical variables by performing a local
conformal transformation of the tetrad:
\begin{equation}
\hat{E}_{\mu }^{a}=\Omega (x)~E_{\mu }^{a},\ \ \ \ \ \ \ \hat{e}_{a}^{\mu }=\Omega ^{-1}(x)~e_{a}^{\mu },
\end{equation}
so it is $\hat{E}=\Omega ^{4}E$ and $\hat{e}=\Omega ^{-4}e$. It is well
known that the action \eqref{actionJ} will transform as \cite{Yang:2010ji,Wright:2016ayu}
\begin{equation}
S=\dfrac{1}{2\kappa }\int d^{4}x~\hat{E}~[\phi ~(\Omega^{-2}~\hat{T}
-4~\Omega^{-3}~\hat{T}^{\mu }~\partial_{\mu }\Omega -6~\Omega^{-4}~\hat{g}
^{\mu \nu }~\partial_{\mu }\Omega ~\partial_{\nu }\Omega )-\Omega
^{-4}~V(\phi )]
\end{equation}
(we can disregard $S_{matter}$ since we are only interested in the number of
gravitational degrees of freedom). We will choose the conformal factor $\Omega $ to have a minimal coupling between $\phi $ and the torsion scalar $T$. The choice $\phi =\Omega ^{2}$ converts the action \eqref{actionJ} into
\begin{equation}
S=\dfrac{1}{2\kappa }\int d^{4}x~\hat{E}~\left[ \hat{T}-2\phi^{-1}~\hat{T}^{\mu }~\partial_{\mu }\phi -\dfrac{3}{2}~\phi^{-2}~\hat{g}^{\mu \nu
}~\partial_{\mu }\phi ~\partial_{\nu }\phi -\phi ^{-2}V(\phi )\right].
\label{actref}
\end{equation}
In this action it would be desirable to have the canonical form for the
kinetic term; therefore it is convenient to redefine the scalar field to a
new field $\psi $ such that
\begin{equation}
\psi =\sqrt{3}~\ln \phi,
\end{equation}
and the action \eqref{actref} is rewritten as
\begin{equation}
S=\dfrac{1}{2\kappa }\int d^{4}x~\hat{E}~\left[ \hat{T}-\dfrac{2}{\sqrt{3}}
~\partial _{\mu }\psi ~\hat{T}^{\mu }-\dfrac{1}{2}~\hat{g}^{\mu \nu
}~\partial _{\mu }\psi ~\partial _{\nu }\psi -U(\psi )\right],
\label{fTeinframe2}
\end{equation}
where the potential is $U(\psi )=\phi ^{-2}V(\phi )$. After an integration
by parts the action reads
\begin{equation}
S=\dfrac{1}{2\kappa }\int d^{4}x~\hat{E}~\left[ \hat{T}+\dfrac{2}{\sqrt{3}}
~\psi ~\hat{e}~\partial _{\mu }(\hat{E}~\hat{T}^{\mu })-\dfrac{1}{2}~\hat{g}
^{\mu \nu }~\partial _{\mu }\psi ~\partial _{\nu }\psi -U(\psi )\right].
\label{fTeinframe1}
\end{equation}
This action exhibits an ordinary teleparallel theory together with a phantom
scalar field. However there is also an annoying term that (nonminimally)
couples the ordinary gravity to the scalar field. So, the Einstein frame in $f(T)$ gravity is unable to cleanly separate the extra degree of freedom, as
it effectively happens in $f(R)$ gravity.

Although we have already shown a rather strong evidence in favor of an extra
degree of freedom in $f(T)$ gravity, the issue can be definitely solved
within the formalism for constrained Hamiltonian systems, as it was done in
a recent publication \cite{Ferraro:2018tpu}. We can gain some insight into
this formalism by looking at how it works in a simple toy model, as we are
going to show in the next section.

\section{\protect\bigskip Modifying a mechanical system with rotational pseudoinvariance}

\label{sec:toymodel}

\subsection{Pseudoinvariant rotational Lagrangian}

Some features of $f(T)$ gravity can be mimicked by deforming the mechanical
Lagrangian
\begin{equation}
L=\frac{1}{2}~\left( \frac{\dot{z}}{z}+\frac{\dot{\overline{z}}}{\overline{z}%
}\right) ^{2}+A~\frac{\dot{z}}{z}+\overline{A}~\frac{\dot{\overline{z}}}{%
\overline{z}}-U(z\overline{z}),  
\label{Lrot}
\end{equation}%
where $z$ and $\overline{z}$ are complex conjugate canonical variables and $A$ and $\overline{A}$ are complex conjugate constants. This Lagrangian can be
rewritten as
\begin{equation}
L=\frac{1}{2}~\left( \frac{d}{dt}\ln (z\overline{z})\right) ^{2}+\frac{d}{dt} 
(A~\ln z+\overline{A}~\ln \overline{z})-U(z\overline{z}),  
\label{Lrot2}
\end{equation}
so it just provides a dynamics to the combination $z\overline{z}$, since the
total derivative term is irrelevant in Lagrange equations. This means that
the evolution of the phase $z/|z|$ is not determined by Lagrange equations.
We can notice this fact also at the level of the symmetries of the
Lagrangian, which is pseudoinvariant under \textit{local} rotations:
\begin{equation}
z \longrightarrow e^{i\alpha (t)}  z,\ \ \ \ \overline{z} \longrightarrow e^{-i\alpha (t)} \overline{z}\ \ \ \Longrightarrow \ \ \ 
L\longrightarrow L+i \dot{\alpha} (A-\overline{A}).
\end{equation}
We can recognize some features that resemble the TEGR theory. In fact, the
TEGR Lagrangian is pseudoinvariant under local Lorentz transformations of
the tetrad;\ so it can only govern the dynamics of the metric, but it is
unable to determine the \textquotedblleft orientation\textquotedblright\ of
the tetrad. The analogy is not complete because the boundary term in this
toy model just contain first derivatives of the canonical variables,
differing from the boundary term in Eq.~\eqref{Lagreq} which is second order
in derivatives of the tetrad.

Now let us pass to the Hamiltonian formalism and look for the constraint
algebra. The canonical momenta are defined as
\begin{equation}
p_{z}\equiv \frac{\partial L}{\partial \dot{z}}=\frac{1}{z}\left( \frac{\dot{z}}{z}+\frac{\dot{\overline{z}}}{\overline{z}}+A\right),\ \ \ \ \ p_{\overline{z}}\equiv \frac{\partial L}{\partial \dot{\overline{z}}}=\frac{1}{\overline{z}}\left( \frac{\dot{z}}{z}+\frac{\dot{\overline{z}}}{\overline{z}}
+\overline{A}\right),
\end{equation}
from which it is easily seen the \textit{primary} constraint
\begin{equation}
G^{(1)}\equiv z\ p_{z}-A-\overline{z}\ p_{\overline{z}} + \overline{A}\approx
 0,
\end{equation}
which fulfills
\begin{equation}
\{G^{(1)},z\overline{z}\} = 0.  
\label{observable}
\end{equation}
The canonical Hamiltonian is
\begin{equation}
H = \dot{z} p_{z} + \dot{\overline{z}} p_{\overline{z}} - L = \frac{1}{2}\left(
\frac{\dot{z}}{z} + \frac{\dot{\overline{z}}}{\overline{z}}\right) ^{2} + U(z\overline{z}) = \frac{1}{8} \left( z\ p_{z}-A+\overline{z}\ p_{\overline{z}}~-
\overline{A}\right)^{2} + U(z\overline{z}),
\end{equation}
while the primary Hamiltonian results
\begin{equation}
H_{p}=\frac{1}{8}~\left( z\ p_{z}-A+\overline{z}\ p_{\overline{z}}-\overline{%
A}\right) ^{2}+U(z\overline{z})+u(t)~G^{(1)},
\end{equation}%
where $u(t)$ is a Lagrange multiplier. We verify that $G^{(1)}$ evolves
without leaving the constraint surface:
\begin{equation}
\dot{G}^{(1)}=\{G^{(1)},H_{p}\}=\frac{1}{4}~(~z\ p_{z}-A+\overline{z}\ p_{%
\overline{z}}-\overline{A})~\{G^{(1)},z\ p_{z}-A+\overline{z}\ p_{\overline{z}}-\overline{A}\}=0.
\end{equation}%
Therefore, no \textit{secondary} constraints appear. There is a unique
\textit{first-class} constraint;\ so one degree of freedom is removed, and
the system is left with just one genuine degree of freedom \cite{Sundermeyer}%
. Equation \eqref{observable} reveals that $z\overline{z}$ is the gauge
invariant (observable) associated to the physical degree of freedom.

\subsection{Modified pseudoinvariant rotational Lagrangian}

Let us deform the theory by replacing the pseudoinvariant Lagrangian with a
function of itself:
\begin{equation}
\mathcal{L}=f(L).
\end{equation}%
This new theory is dynamically equivalent to the one governed by the
Jordan frame Lagrangian that includes an additional dynamical variable $\phi
$:
\begin{equation}
\mathcal{L}=\phi ~L-V(\phi ).  
\label{Jordan}
\end{equation}
In fact, the dynamical equation for $\phi $ is%
\begin{equation}
L~=~V^{\prime }(\phi ).  
\label{eom1}
\end{equation}
So, the dynamics says that $\mathcal{L}$ in Eq.~\eqref{Jordan} is the
Legendre transform of $V(\phi )$; therefore, $\mathcal{L}$ is a function $f(L)$ (each choice of $V$ equals a choice of $f$). The canonical momenta are
\begin{equation}
G_{\pi }^{(1)}\equiv \pi =\frac{\partial \mathcal{L}}{\partial \dot{\phi}}
\approx 0,
\end{equation}
\begin{equation}
p_{z}=\frac{\phi }{z}\left( \frac{\dot{z}}{z}+\frac{\dot{\overline{z}}}{\overline{z}}+A\right),\ \ \ \ \ p_{\overline{z}}=\frac{\phi }{\overline{z}%
}\left( \frac{\dot{z}}{z}+\frac{\dot{\overline{z}}}{\overline{z}}+\overline{A}\right);  
\label{momenta}
\end{equation}
thus one gets the constraint
\begin{equation}
G^{(1)}\equiv z\ p_{z}-A~\phi -\overline{z}\ p_{\overline{z}}+\overline{A}%
~\phi ~\approx ~0.  
\label{G}
\end{equation}
Then the modified system has two noncommuting primary constraints:
\begin{equation}
\{G^{(1)}, G_{\pi }^{(1)}\} =- A + \overline{A}.  
\label{secondclass}
\end{equation}
The canonical Hamiltonian is
\begin{equation}
\mathcal{H}=\dot{z}~p_{z}+\dot{\overline{z}}~p_{\overline{z}}-\mathcal{L}=
\frac{1}{8~\phi }~(~z\ p_{z}-\phi ~A+\overline{z}\ p_{\overline{z}}-\phi ~
\overline{A})^{2}+\phi ~U(z\overline{z})+V(\phi ),
\end{equation}
and the primary Hamiltonian is
\begin{equation}
\mathcal{H}_{p}~=~\mathcal{H}+u^{\pi }(t)~G_{\pi }^{(1)}+u(t)~G^{(1)},
\label{Hamiltonian}
\end{equation}
where $u^{\pi }$ and $u$ are Lagrange multipliers. We must evaluate the
evolution of the primary constraints to look for secondary constraints:
\begin{eqnarray}
\dot{G}_{\pi }^{(1)} &=&\{G_{\pi }^{(1)},\mathcal{H}_{p}\}=\{\pi ,\mathcal{H}_{p}\}  \notag \\
&=&\frac{(z\ p_{z}+\overline{z}\ p_{\overline{z}})^{2}}{8 \phi^{2}}-\frac{1}{8}~(A+\overline{A})^{2}-U(z\overline{z})-V^{\prime }(\phi )~+u(t)~(A-\overline{A})  \notag \\
&=&L-\frac{1}{2}~(A-\overline{A})~\left( \frac{\dot{z}}{z}-\frac{\dot{\overline{z}}}{\overline{z}}\right) - V^{\prime }(\phi )~+~u(t)~(A - \overline{A}).
\end{eqnarray}
\begin{equation}
\dot{G}^{(1)}=\{G^{(1)},\mathcal{H}_{p}\}=-u^{\pi }~(A-\overline{A}).
\end{equation}
Thus, the consistency of the constraints with the evolution can be obtained
by choosing the Lagrange multipliers as%
\begin{equation}
u^{\pi } = 0,\ \ \  u(t)=\frac{L(t)-V^{\prime }(\phi (t))}{A-\overline{A}}~-\frac{1}{2}\left( \frac{\dot{z}(t)}{z(t)}-\frac{\dot{\overline{z}}(t)}{\overline{z}(t)}\right).  
\label{multipliers}
\end{equation}
Therefore the system has no secondary constraints; the only constraints $G^{(1)}$ and $~G_{\pi }^{(1)}$ are second-class [see Eq.~\eqref{secondclass}].
So, they remove only one degree of freedom \cite{Sundermeyer}; there are two
genuine degrees of freedom among the variables $(z,\overline{z},\phi )$.
Notice that, differing from $f(T)$ gravity, no gauge freedom is left in this
system since both Lagrange multipliers have been fixed; so the primary
Hamiltonian completely determines the evolution of the variables. In
particular, the evolution of $\phi $ is frozen:
\begin{equation}
\dot{\phi} = \{\phi ,\mathcal{H}_{p}\} = u^{\pi } = 0.
\end{equation}
Coming back to the Lagrange equations, let us use for $L$ the form 
\eqref{Lrot2}; then by varying $\mathcal{L}$ with respect to $z\overline{z}$ and $z$ one obtains
\begin{equation}
\frac{d}{dt}\left( \phi \frac{d}{dt}\ln(z\overline{z})\right) = -\phi U^{\prime }(z\overline{z}),\ \ \ \  \frac{d}{dt}\left( \phi
 A\right) = 0.
\end{equation}
So, we retrieve the result that $\phi $ is constant. Besides $z\overline{z}$
obeys the same dynamical equation of the nondeformed theory. In particular the magnitude
\begin{equation}
H =\frac{1}{2}~\left( \frac{d}{dt}\ln (z\overline{z})\right) ^{2}+U(z \overline{z})  \label{H}
\end{equation}
is a constant of motion. Since $\phi $ is constant, then the dynamical
equation for $\phi $, i.e. Eq.~\eqref{eom1}, implies that $L$ is constant.
From Eqs.~\eqref{eom1}, \eqref{Lrot} and \eqref{H} we obtain that the
evolution of the phase of $z$ is determined by the dynamics through the
first-order equation
\begin{equation}
\frac{d}{dt}(A~\ln z+\overline{A}~\ln \overline{z})=-\frac{1}{2}~\left(
\frac{d}{dt}\ln (z\overline{z})\right) ^{2}+U(z\overline{z})+V^{\prime
}(\phi )=2~U(z\overline{z})+V^{\prime }(\phi )-H.  \label{phase}
\end{equation}
At the initial conditions, the choice of $\phi $ is the way one chooses the
initial velocity of the phase. Due to the remaining \textit{global}
rotational symmetry, the initial value of the phase is irrelevant. This
equation can be also obtained in the Hamiltonian formalism by computing the
Poisson bracket between $A~\ln z+\overline{A}~\ln \overline{z}$ and $\mathcal{H}_{p}$, and then substituting the momenta and the Lagrange
multipliers with \eqref{momenta} and \eqref{multipliers}.

In sum, the system described by the Lagrangian \eqref{Jordan} has two
degrees of freedom:\ one of them is $|z|^{2}=z\overline{z}$ whose dynamics
does not differ from the one described by the Lagrangian \eqref{Lrot2}; in
both cases we arrive to the conserved quantity $H$ in Eq.~\eqref{H}. Once
the evolution of $|z|$ is determined by the choices of the initial value $%
|z(t_{o})|$ and the constant of motion $H$, Eq.~\eqref{phase} determines the
evolution of the phase of $z$, which is the remaining degree of freedom. In
this equation, the value of the constant $\phi $ is connected to the initial
value of the derivative of the phase of $z$; there is no other physics
associated with $\phi $ over and above the one related to the phase of $z$.

In analogy with $f(T)$ gravity, $\phi $ could be then regarded as a
variable carrying information about the orientation  of the tetrad, which would be partially
determined by the dynamical equations.

\section{Hamiltonian formalism and frame dependence}

\label{sec:HfT}

The Hamiltonian formalism for $f(T)$ gravity, starting from the
(scalar-torsion) Jordan frame formulation of the theory, was presented in
full detail in \cite{Ferraro:2018tpu}. There are several unusual features of
this physical system that deserves a comment. In most Hamiltonian
constrained systems the identification of secondary constraints is a rather
trivial enterprise, and there is no need to calculate the left and right
null eigenvectors of the matrix of Poisson brackets among constraints.
However, in the Hamiltonian formalism for $f(T)$ gravity we have to pay
close attention to this point, which makes the theory an atypical example of
a constrained Hamiltonian system.

After following the Dirac-Bergmann algorithm for $f(T)$ gravity one is led
to a set of first-class $\Phi _{I}$ and second-class constraints $\chi _{A}$
organized in the following way (see definitions at \cite{Ferraro:2018tpu}):

\begin{itemize}
\item $n$ first-class constraints $\Phi _{c}=G_{c}^{(1)}$ coming from the
absence of $\dot{E}_{0}^{c}$ in the Lagrangian,

\item $n-1$ first-class constraints $\Phi _{i}=G_{i}^{(2)}$ associated with
the spatial diffeomorphisms (supermomentum constraints),

\item $\frac{n(n-1)}{2}-1$ first-class constraints $\Phi _{ab}=\tilde{G}
_{ab}^{(1)}$ associated with the Lorentz algebra,

\item one first-class constraint $\Phi _{0}=\tilde{G}_{0}^{(2)}$associated
with the temporal diffeomorphism (super-Hamiltonian constraint), and

\item two second-class constraints: $\chi _{\pi }=G_{\pi }^{(1)}$, coming
from the absence of $\dot{\phi}$ in the Lagrangian, and $\chi =\alpha^{ab} \tilde{G}_{ab}^{(1)}$ associated with the Lorentz algebra.
\end{itemize}

This classification is obtained after a redefinition of the original Lorentz
constraints that maximizes the number of vanishing Poisson brackets in the
algebra of constraints. In fact, in terms of the original constraints, the
nonvanishing Poisson brackets involve the Lorentz constraints and the
super-Hamiltonian constraint; they are
\begin{eqnarray}
\{G_{ab}^{(1)},G_{\pi }^{(1)}\} &\equiv &F_{ab},  \notag \\
\{G_{0}^{(2)},G_{\pi }^{(1)}\} &\equiv &F_{\phi }.  \label{Poisson}
\end{eqnarray}
Since $F_{ab}$ and $F_{\phi }$ are functions of the canonical coordinates and
momenta, they could vanish at particular points of the phase space. So, let
us suppose that $F_{01}$ is not zero in some neighborhood of the constraint
surface. Then we can change the basis of the constraint algebra by combining
the original Lorentz constraints $G_{ab}^{(1)}$ in the following way:
\begin{eqnarray}
\tilde{G}_{01}^{(1)} &=&G_{01}^{(1)},  \notag \\
\tilde{G}_{02}^{(1)} &=&F_{01}~G_{02}^{(1)}-F_{02}~G_{01}^{(1)},  \notag \\
&&\vdots  \notag \\
\tilde{G}_{(n-2)(n-1)}^{(1)}
&=&F_{01}~G_{(n-2)(n-1)}^{(1)}-F_{(n-2)(n-1)}~G_{01}^{(1)}.  \label{LorFC}
\end{eqnarray}
Besides, we replace the super-Hamiltonian constraint $G_{0}^{(2)}$ with
\begin{equation}
\tilde{G}_{0}^{(2)}=F_{01}~G_{0}^{(2)}-F_{\phi }~G_{01}^{(1)}.
\label{newHconstr}
\end{equation}
Thus the only nonvanishing Poisson brackets will be
\begin{equation}
\{\tilde{G}_{01}^{(1)}, G_{\pi }^{(1)}\} \equiv F_{01}.
\end{equation}
In this way we have two second-class constraints $\chi _{A}$, the remaining
constraints being first-class. The matrix $\Delta _{AB}=\{\chi _{A},\chi
_{B}\}$,
\begin{equation}
\Delta _{AB}=\left(
\begin{array}{cc}
\{G_{\pi }^{(1)},G_{\pi }^{(1)}\} & \{\tilde{G}_{01}^{(1)},G_{\pi }^{(1)}\}
\\
\{G_{\pi }^{(1)},\tilde{G}_{01}^{(1)}\} & \{\tilde{G}_{01}^{(1)},\tilde{G}
_{01}^{(1)}\}
\end{array}
\right) =\left(
\begin{array}{cc}
0 & F_{01} \\
-F_{01} & 0
\end{array}
\right),
\end{equation}
is a $2\times 2$ invertible matrix. The choice that $\tilde{G}_{01}^{(1)}$
be the odd second-class Lorentz constraint is completely arbitrary and any
linear combination of the form
\begin{equation}
\chi =\alpha^{ab} G_{ab}^{(1)}
\end{equation}
can be chosen to be the second-class one (but $\alpha^{ab} F_{ab}$ must be
different from zero). This is because the Dirac-Bergmann algorithm does not
determine a particular linear combination to be second-class. However the
arbitrariness of the choice does not affect the conclusions.

The equations assuring the consistency of the evolution of the first-class
constraints result to be trivial, while the ones for the second-class
constraints $\chi _{A}=(G_{\pi }^{(1)},\tilde{G}_{01}^{(1)})$,
\begin{equation}
\{\chi_{A}, H_{c}\} + \Delta_{AB} u^{B} \overset{!}{\approx }0,
\end{equation}
can be satisfied by properly choosing the Lagrange multipliers $u^{\pi }$
and $\widetilde{u}^{01}$:
\begin{equation}
u^{\pi }=0, \ \ \ \ \ \ F_{01} \widetilde{u}^{01} = F_{\phi },
\end{equation}
which is comparable to Eq.~\eqref{multipliers} in Sec.~\ref{sec:toymodel}. Therefore, when passing from TEGR to $f(T)$ gravity as
described in the Jordan frame, one of the first-class constraints associated
with the Lorentz algebra becomes a second-class constraint paired with $G_{\pi }^{(1)}$. As was shown in the toy model of the previous section, this
fact amounts to the retrieval of a degree of freedom that took part in the
gauge freedom of the TEGR theory. Concretely, the invariance under local
Lorentz transformations is partially lost since a linear combination of
their generators is no longer a gauge transformation. In other words, some
feature of the tetrad orientation has passed to be governed by the dynamics and now forms part of the set of genuine degrees of freedom together with
the metric degrees of freedom.

For defining the set of first-class Lorentz constraints \eqref{LorFC} we had
to assume that at least one of the $F_{ab}$ is nonvanishing. However, there
could be points on the constraint surface where all the $F_{ab}$ vanishes.
In this case, the consistency of the Dirac-Bergmann algorithm could impose
two different outcomes: $F_{\phi}$ also vanishes identically and the theory
has the same d.o.f. as TEGR, or $F_{\phi}$ is a constraint and the
algorithm continues. Although, for obtaining a meaningful number of d.o.f.,
the algorithm is not allowed to extend too much, as we will argue next. Let
us reasonably assume that neither $G^{(2)}_{\mu}$ nor $G^{(1)}_c$ become
second-class and thus they remove $2n$ d.o.f. But $X$ of the $n(n-1)/2$
Lorentz constraints could still become second-class due to the pairing with
the same amount of second-class constraints. Besides, there could appear
additional $Y_1$ first-class constraints (f.c.c.) and $Y_2$ second-class
constraints (s.c.c.). The counting of d.o.f. would go in the following way:
\begin{equation}
\begin{split}
\text{number of d.o.f.} & = n^2 + 1 - (\text{number of f.c.c.}) - \frac{1}{2}
(\text{number of s.c.c.} ) \\
& = n^2 + 1 - (2n + \frac{n}{2}(n-1)-X + Y_1 ) - \frac{1}{2}(2X + Y_2 ) \\
& = \frac{n(n-3)}{2}+1-Y_1 - \frac{1}{2} Y_2.
\end{split}
\label{naive_dof}
\end{equation}
The case of $f(T)$ gravity falls in the category $X=1$, since there is only
one constraint $G^{(1)}_{\pi}$ that pairs up with one Lorentz constraint to
become second-class. If we are in a particular point of the phase space
where all the $F_{ab}$ are zero, the Poisson bracket among $G^{(1)}_{\pi}$
and $G^{(1)}_{ab}$ would vanish on shell, but some of the $G^{(1)}_{ab}$
could still become second-class due to the appearance of new constraints
represented by $X$. In this case the algorithm would require $F_{\phi}$ to
weakly vanish on shell and become a constraint. If it were first-class, then
$Y_1=1$, $Y_2=0$; thus Eq.~\eqref{naive_dof} would reproduce the number
of d.o.f. of TEGR. Instead, if both $F_{\phi}$ and $G^{(1)}_{\pi}$ are
second-class and they originate at least one new first- or two second-class
constraints, the counting of d.o.f. will give a meaningless answer: less
d.o.f. than TEGR!. This, of course, is an unacceptable result, and the only
reasonable possibility is that the pair $F_{\phi},G^{(1)}_{\pi}$ is
second-class and removes one d.o.f. to give the TEGR result. In this case,
the time stability of $F_{\phi}$ should be proved calculating its Poisson
algebra with the remaining constraints; however it is a very involved
calculation which may require an improved Hamiltonian formalism.

Finally, it is worth mentioning a puzzling feature in the Hamiltonian
constraint of Eq.~\eqref{newHconstr}, since the second term of this
expression may have a cubic dependence on the canonical momenta. This is
easily seen because $G_{01}^{(1)}$ is linear in the momenta, but $F_{\phi }$
depends on the torsion scalar and is therefore quadratic in the momenta. The
cubic dependence of $\tilde{G}_{0}^{(2)}$ is the price to be paid for
getting a first-class Hamiltonian constraint. This fact can be traced to the
pseudoinvariance of the TEGR Lagrangian, which implies that the $f(T)$
action breaks the local Lorentz invariance. This is an unusual feature for a
Hamiltonian system, and it could lead to instabilities since the Hamiltonian
could be unbounded from below \cite{Ostrogradsky:1850,Woodard:2015zca}.

\section{Hamiltonian approach in the teleparallel Einstein frame}

\label{sec:HTEF}

Since the conformal transformation that leads from the Jordan frame to the
Einstein frame is a canonical transformation, one should expect the same
number of d.o.f.~in both the Jordan and Einstein frames. Nonetheless this
mere verification could throw light on the issue of the meaning of the extra
d.o.f. However this exercise will indeed reveal the intricate Hamiltonian
structure of $f(T)$ gravity in the Einstein frame. Let us start from a
gravitational theory whose action is
\begin{equation}
S[E_{\mu }^{a},\psi ]=\dfrac{1}{2\kappa }\int d^{4}x\ E~\left( T+\dfrac{2}{%
\sqrt{3}}\psi B-\dfrac{1}{2}g^{\mu \nu }~\nabla _{\mu }\psi ~\nabla _{\nu
}\psi +U(\psi )\right).  
\label{EinstOstro}
\end{equation}
We get rid of the hatted fields in the action and start working directly
under the assumption that the teleparallel Einstein frame is our theory for
describing the gravitational phenomena. In this formulation it is simple to see that the Hamiltonian procedure is not applicable, since the term $B$, as we have written it in Eq.\eqref{Bterm}, has an explicit dependence on second-order time derivatives on the tetrad field. In this case we need to resort to Ostrogradsky's choice for canonical variables and momenta. It is well known that Lagrangians having higher-order time derivatives might suffer from the Ostrogradsky instability \cite{Ostrogradsky:1850,Woodard:2015zca}, although as we will prove later, in our case we can get rid of this obstruction by means of an integration by parts. However, motivated by a deeper understanding in this subject, it will be interesting to first explore this approach. Ostrogradsky's definitions of canonical momenta are given by
\begin{equation}
\Pi_{a}^{\mu } = \dfrac{\partial L}{\partial (\partial_{0}E_{\mu }^{a})}
-\partial_{0} \dfrac{\partial L}{\partial (\partial _{0}\partial_{0} E_{\mu}^{a})}, \ \ \ \ \  \mathcal{P}_{a}^{\mu } = \dfrac{\partial L}{\partial (\partial_{0} \partial_{0} E_{\mu }^{a})},  \label{canOstro}
\end{equation}
that are associated to the canonical variables $E_{\mu }^{a}$ and $\mathcal{E}_{\mu }^{a} \equiv \partial_{0}E_{\mu }^{a}$, respectively. Within the
premetric formalism developed in Sec.~\ref{sec:tegrfT}, they can be
easily computed; the result is
\begin{equation}
\Pi_{a}^{\mu } = E \partial_{\rho } E_{\lambda }^{b}~e_{c}^{0}~e_{e}^{\mu
}~e_{d}^{\rho }~e_{f}^{\lambda }~\left( M_{ab}^{\ \ cedf}+\dfrac{2}{\sqrt{3}}
~\psi ~B_{ab}^{\ \ cedf}\right) -\dfrac{2}{\sqrt{3}}~\partial _{0}[\psi
~E~e_{b}^{0}~e_{c}^{\mu }~e_{d}^{0}]~(\delta _{a}^{d}~\eta ^{bc}-\delta
_{a}^{c}~\eta ^{bd})  
\label{Pi1EF}
\end{equation}
and
\begin{equation}
\mathcal{P}_{a}^{\mu }=\dfrac{2}{\sqrt{3}}~\psi ~E~e_{b}^{0}~e_{c}^{\mu
}~e_{d}^{0}\ (\delta _{a}^{d}~\eta ^{bc}-\delta _{a}^{c}~\eta ^{bd}).
\label{Pi2EF}
\end{equation}
There is also a canonical momentum associated with the scalar field $\psi $,
given by
\begin{equation}
\pi =\dfrac{\partial L}{\partial (\partial _{0}\psi )}=-E~g^{0\nu }~\partial
_{\nu }\psi.  
\label{pi}
\end{equation}
It is clear that there appear $n$ primary constraints $\mathcal{P}_{a}^{0}\approx 0$. Besides, \eqref{Pi1EF} can be rewritten as a primary
constraint using \eqref{pi}, which is
\begin{equation}
\Pi _{a}^{\mu }=E\ \mathcal{E}_{\lambda }^{b}\ e_{e}^{\mu }~e_{f}^{\lambda }~%
\tilde{C}_{ab}^{\ \ ef}+E~\partial _{i}E_{\lambda }^{b}~e_{c}^{0}~e_{e}^{\mu
}~e_{d}^{i}~e_{f}^{\lambda }~\left( M_{ab}^{\ \ cedf}+\dfrac{2}{\sqrt{3}}%
\psi ~B_{ab}^{\ \ cedf}\right) +\dfrac{2}{\sqrt{3}}\left( \dfrac{e}{g^{00}}%
~\pi +\dfrac{g^{00}}{g^{0i}}~\partial _{i}\psi \right) E[e_{a}^{0}~g^{0\mu
}-e_{a}^{\mu }~g^{00}],
\end{equation}
where
\begin{equation}
\tilde{C}_{ab}^{\ \ ef}=e_{c}^{0}~e_{d}^{0}\left( M_{ab}^{\ \ cedf}+\dfrac{2
}{\sqrt{3}}\psi ~B_{ab}^{\ \ cedf}-\dfrac{4}{\sqrt{3}}~\psi ~(\eta
^{cf}~\delta _{a}^{[e}\delta _{b}^{d]}+\eta ^{cd}~\delta _{a}^{[f}\delta
_{b}^{e]}+\eta ^{ec}~\delta _{a}^{[d}\delta _{b}^{f]}+\delta _{b}^{d}~\eta
^{f[c}\delta _{a}^{e]})\right).
\end{equation}
As depicted in \cite{Woodard:2015zca}, Eqs.~\eqref{Pi1EF}, \eqref{Pi2EF} and \eqref{pi} should be used to write the canonical
Hamiltonian
\begin{equation}
H_{c}=\Pi _{a}^{\mu }~\mathcal{E}_{\mu }^{a}+\mathcal{P}_{a}^{\mu }~\partial
_{0}\mathcal{E}_{\mu }^{a}+\pi ~\partial _{0}\psi -L  \label{EFHc}
\end{equation}
in terms of the canonical variables $(E_{\mu }^{a},\Pi _{a}^{\mu },\mathcal{E}_{\mu }^{a},\mathcal{P}_{a}^{\mu },\psi ,\pi )$ and their spatial
derivatives. According to the formalism, the acceleration $\mathcal{A}_{\mu
}^{a}=\partial _{0}\partial _{0}E_{\mu }^{a}$ should be solved as a function
of the canonical variables from the second equation in Eq.~\eqref{canOstro};
thus, $\Pi _{a}^{\mu }$ will not enter in the expression for $\mathcal{A}%
_{\mu }^{a}$. Therefore, $H_{c}$ will depend on $\Pi _{a}^{\mu }$ only
through the first term of Eq.~\eqref{EFHc}, which is a linear in $\Pi
_{a}^{\mu }$ (in this higher-order formalism, $\mathcal{E}_{\mu }^{a}$ is a
canonical variable). This means that $H_{c}$ is unbounded from below and we might be in the presence of some sort of instability, although briefly we will see that we can circumvent this issue resorting to an equivalent representation.

However, the Lagrangian in Eq.~\eqref{EinstOstro} is linear in the
acceleration $\mathcal{A}_{\mu }^{a}=\partial _{0}\partial _{0}E_{\mu
}^{a}=\partial _{0}\mathcal{E}_{\mu }^{a}$ . Therefore the definition of $\mathcal{P}_{a}^{\mu }$ does not allow one to solve the acceleration in terms of
the canonical variables but leads to a constraint [cf. Eq.~\eqref{Pi2EF}].
Thus, the acceleration will remain as a Lagrange multiplier in the canonical
Hamiltonian \eqref{EFHc}, which will be subsumed in the respective Lagrange
multiplier of the primary Hamiltonian.

Ostrogradsky's procedure can be avoided by performing an integration by
parts in \eqref{EinstOstro}; thus we will work with the action
\begin{equation}
S[E_{\mu }^{a},\psi ]=\dfrac{1}{2\kappa }\int d^{4}x~E~\left( T-\dfrac{2}{\sqrt{3}}~T^{\mu }~\partial _{\mu }\psi -\dfrac{1}{2}~g^{\mu \nu }~\nabla
_{\mu }\psi ~\nabla_{\nu }\psi + U(\psi )\right).  \label{SEinstdPhi}
\end{equation}
In this case, the canonical momenta associated with the tetrad field results
\begin{equation}
\Pi _{a}^{\mu }=E~\partial _{\rho }E_{\lambda }^{b}~e_{c}^{0}~e_{e}^{\mu
}~e_{d}^{\rho }~e_{f}^{\lambda }~M_{ab}^{\ \ cedf}-\dfrac{2}{\sqrt{3}}
~E~\partial _{\nu }\psi ~e_{b}^{0}~e_{c}^{\nu }~e_{d}^{\mu }~[\eta
^{dc}\delta _{a}^{b}-\eta ^{bc}\delta _{a}^{d}],  \label{piE}
\end{equation}
while the canonical momentum for the scalar field is
\begin{equation}
\pi =-E~g^{0\nu }~\partial _{\nu }\psi -\dfrac{2}{\sqrt{3}}~E~\partial
_{\rho }E_{\nu }^{a}~e_{b}^{0}~e_{c}^{\rho }~e_{d}^{\nu }~[\eta ^{bd}\delta
_{a}^{c}-\eta ^{cb}\delta _{a}^{d}].  
\label{pipsi}
\end{equation}
It is easy to see that both canonical momenta include canonical velocities $%
\partial _{0}E_{\mu }^{a}$ and $\partial _{0}\psi $ in their definitions.
Therefore, the procedure to write the canonical Hamiltonian in terms of the
canonical variables will be more intricate than the one developed for TEGR
in Ref. \cite{Ferraro:2016wht}. Let us solve for $\partial _{0}\psi $ in \eqref{pipsi}:
\begin{equation}
\partial _{0}\psi =-\dfrac{e}{g^{00}}~\pi -\dfrac{g^{0i}}{g^{00}}~\partial
_{i}\psi -\dfrac{2}{\sqrt{3}g^{00}}~\partial _{\rho }E_{\lambda
}^{a}~e_{b}^{0}~e_{c}^{\rho }~e_{d}^{\lambda }~[\eta ^{bd}\delta
_{a}^{c}-\eta ^{cb}\delta _{a}^{d}].
\end{equation}
After replacing this result in \eqref{piE} and performing some
arrangements, the following expression is obtained:
\begin{eqnarray}
E_{\mu }^{e}~\Pi _{a}^{\mu } &=&E~\partial _{0}E_{\lambda
}^{b}~e_{f}^{\lambda }~\tilde{C}_{ab}^{\ \ ef}+E~\partial _{i}E_{\lambda
}^{b}~e_{f}^{\lambda }~\left( e_{c}^{0}~e_{d}^{i}~M_{ab}^{\ \ cedf} + \dfrac{%
4}{3g^{00}}~e_{c}^{0}~e_{d}^{0}~e_{g}^{0}~e_{h}^{i}~[\eta ^{ce}\delta
_{a}^{d}-\eta ^{cd}\delta _{a}^{e}][\eta ^{gf}\delta _{b}^{h}-\eta
^{hg}\delta _{b}^{f}]\right)  \notag \\
&&+\dfrac{2~\pi }{\sqrt{3}~g^{00}}~e_{c}^{0}~e_{d}^{0}~[\eta^{ce}\delta
_{a}^{d}-\eta ^{cd}\delta _{a}^{e}]+\dfrac{2}{\sqrt{3}}~E~\partial _{i}\psi
~e_{d}^{0}~\left( -e_{c}^{i}+e_{c}^{0}~\dfrac{g^{0i}}{g^{00}}\right) ~[\eta
^{ce}\delta _{a}^{d}-\eta ^{cd}\delta _{a}^{e}],  \label{Piea}
\end{eqnarray}
where
\begin{equation}
\tilde{C}_{ab}^{\ \ ef}=e_{c}^{0}~e_{d}^{0}~M_{ab}^{\ \ cedf}+\dfrac{4}{3g^{00}}~e_{d}^{0}~e_{c}^{0}~e_{g}^{0}~e_{h}^{0}~[\eta ^{ce}~\delta
_{a}^{d}-\eta ^{cd}~\delta _{a}^{e}][\eta ^{gf}~\delta _{b}^{h}-\eta
^{hg}~\delta _{b}^{f}].  
\label{CEinFrame}
\end{equation}
Apart from looking for primary constraints, the next obvious step would be
to write the canonical Hamiltonian associated with \eqref{SEinstdPhi}. It
can be attempted by means of the Moore-Penrose pseudoinverse method
introduced in \cite{Ferraro:2016wht} in order to invert the matrix \eqref{CEinFrame}. For this, it would be useful to know whether its
eigenvalues and eigenvectors are the same obtained for $C_{ab}^{\ \ ef}$ in
TEGR; however the answer is negative. Moreover, the structure of eigenvalues
and eigenvectors is more complicated, and it may require to develop advanced
mathematical techniques in order to find them. The fact that the matrix %
\eqref{CEinFrame} is not the same than the one in TEGR is intriguing and
puts forward the question whether the teleparallel Einstein frame can
genuinely be interpreted as TEGR plus a scalar field. The answer to this
interesting question calls for further research.

Finally we can attempt to obtain the primary constraints of this theory by
multiplying Eq.~\eqref{Piea} by the eigenvectors $v_{|g|e}^{\ \ a} = e_{e}^{0} \delta_{g}^{a}$ and $v_{|gh|e}^{\ \ \ a} = \delta_{g}^{a} \eta_{he} - \delta_{h}^{a} \eta_{ge}$. The first one gives the trivial constraint
\begin{equation}
G_{g}^{(1)}=\Pi_{g}^{0}\approx 0.
\end{equation}
For the second set of eigenvectors $v_{|gh|e}^{\ \ \ a}$, it is easy to
check that they are null eigenvectors of $\tilde{C}_{ab}^{\ \ ef}$, i.e.
they satisfy $v_{|gh|e}^{\ \ \ a}\tilde{C}_{ab}^{\ \ ef}=0$, as they are
also for the matrix $C_{ab}^{\ \ ef}$ in TEGR \cite{Ferraro:2016wht}. Using
this result in \eqref{Piea}, the following constraints are obtained:
\begin{equation}
G_{ab}^{(1)}=2~\eta _{e[b}~\Pi _{a]}^{i}~E_{i}^{e}+4~E~\partial
_{i}E_{j}^{b}~e_{[b}^{0}~e_{a}^{i}~e_{c]}^{j}+\dfrac{2}{\sqrt{3}}~\partial
_{i}\psi ~(e_{a}^{i}~e_{b}^{0}-e_{a}^{0}~e_{b}^{i}).
\end{equation}
The appearance of an extra term with respect to the TEGR case changes
drastically the Lorentz algebra, as this term has a  nonvanishing Poisson
bracket with the $G_{cd}^{(1)}$'s. It is not difficult to prove that
\begin{equation}
\{G_{ab}^{(1)},G_{cd}^{(1)}\}=\eta _{bd}~G_{ac}^{(1)}+\eta
_{ac}~G_{bd}^{(1)}-\eta _{bc}~G_{ad}^{(1)}-\eta _{ad}~G_{bc}^{(1)}+\dfrac{4}{%
\sqrt{3}}\partial _{i}\psi ~\left[ H_{ab}~H_{cd}^{i}-H_{cd}~H_{ab}^{i}\right],  
\label{nonLorentz}
\end{equation}
where
\begin{equation}
H_{ab} \equiv e_{a}^{0} e_{0b} - e_{b}^{0} e_{0a}, \ \ \ H_{cd}^{i} \equiv e_{c}^{0} e_{d}^{i} - e_{d}^{0} e_{c}^{i}.
\end{equation}
These two terms reflect the departure from Lorentz invariance, since they do not vanish in the most general case. However, unlike the teleparallel Jordan frame case, here the theory seems to present full Lorentz violation. That is, none of the Lorentz constraints seems to be first-class, then one might be tempted to conclude that this leads to the counting of d.o.f. of \cite{Li:2011rn}. Nevertheless, there is still another Poisson bracket $\{ G^{(1)}_g, G^{(1)}_{ab} \}$ to be taken into account, which gives
\begin{equation}
\{ G^{(1)}_g, G^{(1)}_{ab} \} = \dfrac{2}{\sqrt{3}} \partial_i \psi e^0_g (e^i_a e^0_b - e^0_a e^i_b). 
\label{noncommgab}
\end{equation}
The noncommuting character of $G^{(1)}_g$ was expected, since gauge transformations of the tetrad in the form $\delta E^a_0 = \epsilon^a$ are no longer symmetries of the Lagrangian due to the dependence on $\partial_0 E^a_0$ in the coupling term proportional to $T^{\mu}\partial_{\mu }\psi$ in the action \eqref{SEinstdPhi} (see Sec. V of \cite{Ferraro:2016wht}). Henceforth, the counting of d.o.f. in this frame requires a rigorous analysis of the time evolution of all constraints with the primary Hamiltonian, a point that needs to be investigated in future work.

\section{Conclusions}

\label{sec:concl}

We have analyzed the issue of the d.o.f.~in $f(T)$ gravity from several
perspectives. For the sake of a comparison, we first  reviewed the
interpretation of the additional d.o.f.~in $f(R)$ gravity in both Jordan and
Einstein frames. Analogously, the study of the equations of motion and its
trace in $f(T)$ gravity reveals evidence of a unique additional d.o.f.~in
the teleparallel Jordan and Einstein frames. To mimic the essential features
of $f(T)$ gravity, we have exhibited the Hamiltonian dynamics of a simple
toy model with pseudorotational invariance. The nonlinear deformation of
this toy model shows the appearance of a scalar degree of freedom $\phi $,
whose value is connected to the initial value of the derivative of the phase
of the complex canonical variable $z$. Analogously, in $f(T)$ gravity the
scalar d.o.f. is connected with the proper parallelization of the spacetime.

Concerning our previous work on the extra d.o.f. of $f(T)$ gravity \cite{Ferraro:2018tpu}, we have emphasized that such analysis can depend on the
point of the phase space under consideration. In fact, the Poisson brackets
between constraints are functions of the tetrad field and their canonical
momenta. Thus, there could exist a point or a neighborhood of the constraint
surface where all the $F_{ab}$'s in Eq.~\eqref{Poisson} were zero, which 
would dramatically change the counting of d.o.f. In that particular points
the consistency of our analysis would imply that $F_{\phi}$ becomes a
constraint, nonetheless if its time consistency does not generate additional
constraints, the theory would have the same number of d.o.f. as TEGR. This
could be suggesting that only some reference frames do manifest the extra
d.o.f.; however it is still unclear the conditions that they should
accomplish.

Finally, we have just introduced two possible ways of applying the
Hamiltonian formalism in the teleparallel Einstein frame. In spite of
appearances, the Einstein frame proves to be rather more involved than the
Jordan frame. This is because the term coupling the vector part of the
torsion $T^{\mu }$ to the scalar field produces an intricate binding of the
canonical momenta, which result in serious difficulties for calculating the
Hamiltonian.

It is clear that further research needs to be done in order to understand
the implications of the additional d.o.f.~of the theory. A simple strategy
to identify it would be to resort to solutions to $f(T)$ gravity in $2+1$
dimensions, where neither TEGR nor GR possess genuine degrees of freedom.
This issue will be addressed in future work \cite{Bejarano:2018}.

\section*{Acknowledgments}

The authors are indebted to C. Bejarano and F. Fiorini for many discussions
and constant support. M.J.G. would like to thank for feedback and fruitful
discussions D. Blixt, M. Hohmann, T. Koivisto, M. Kr\v{s}\v{s}\'{a}k, S. P\'{e}rez-Bergliaffa, C. Pfeifer and E. N. Saridakis, and especially to A.
Golovnev for constructive criticism. M.J.G.  acknowledges the hospitality of the
organizers of the Teleparallel Gravity Worskhop in Tartu 2018, where part of
this work was completed, and to the participants of the workshop for helpful
discussion on the results of \cite{Ferraro:2018tpu}. M.J.G. was partially
supported by ICTP. The work of both authors was supported by Consejo
Nacional de Investigaciones Cient\'{\i}ficas y T\'{e}cnicas (CONICET) and
Universidad de Buenos Aires. R.F. is a member of Carrera del Investigador Cient\'{\i}fico (CONICET, Argentina).

\end{document}